\documentclass[12pt]{iopart}

\usepackage{graphicx}
\usepackage{setstack}

\begin{document}
\title[The odd-mass nuclei in IVBM]{Collective states of the odd-mass nuclei
within the framework of the \\ Interacting Vector Boson Model}

\author{H. G. Ganev}
\address{Institute for Nuclear Research and Nuclear Energy,\\
Bulgarian Academy of Sciences, Sofia 1784, Bulgaria}
\ead{huben@inrne.bas.bg}

\begin{abstract}

A supersymmetric extension of the dynamical symmetry group
$Sp^{B}(12,R)$ of the Interacting Vector Boson Model (IVBM), to the
orthosymplectic group $OSp(2\Omega/12,R)$ is developed in order to
incorporate fermion degrees of freedom into the nuclear dynamics and
to encompass the treatment of odd mass nuclei. The bosonic sector of
the supergroup is used to describe the complex collective spectra of
the neighboring even-even nuclei and is considered as a core
structure of the odd nucleus. The fermionic sector is represented by
the fermion spin group $SO^{F}(2\Omega)\supset SU^{F}(2)$.

The so obtained, new exactly solvable limiting case is applied for
the description of the nuclear collective spectra of odd mass
nuclei. The theoretical predictions for different collective bands
in three odd mass nuclei, namely $^{157}Gd$, $^{173}Yb$ and
$^{163}Dy$ from rare earth region are compared with the experiment.
The $B(E2)$ transition probabilities for the $^{157}Gd$ and
$^{163}Dy$ between the states of the ground band are also studied.
The important role of the symplectic structure of the model for the
proper reproduction of the $B(E2)$ behavior is revealed. The
obtained results reveal the applicability of the models extension.

\end{abstract}

PACS {21.60.Fw, 21.60.Ev, 21.10.Re, 27.70.+q, 23.20.-g}
\maketitle

\section{Introduction}

Symmetry is an important concept in nuclear physics. In finite
many-body systems of this type, it appears as time reversal, parity,
and rotational invariance, but also in the form of dynamical
symmetries.

Many collective properties of the nuclei have been investigated
using models based on dynamical groups. One of the most popular and
widely used models of this type are the Interacting Boson Model
(IBM) \cite{IBM} and its extensions \cite{IS},\cite{IBMext} as well
as the symplectic model \cite{RR} based on the group $Sp(6,R)$. In
these, algebraic models the bands of collective states are
classified by the irreducible representations (irreps) of the
corresponding chains of groups and their corresponding properties,
such as energy levels and electromagnetic transition strengths, are
determined by algebraic methods.

It is well known that nucleons have intrinsic spin and that there
are strong spin-orbit interactions. Moreover, the experiment
reveals, that the presence of spin does not prevent the appearance
of rotational bands. It also establishes similarity in the
rotational character of the different collective bands for
neighboring even-even and odd-even nuclei far from closed shells.
For the description of the nuclear spectra of such even-even nuclei
the above mentioned variety of boson models is used. This is
possible because in the even-even nuclei the pairs of nucleons are
usually considered as coupled to integer angular momentum. However,
this is not the case for odd mass nuclei. Thus, the following
question naturally arises:\ how to incorporate fermion degrees of
freedom into the nuclear dynamics in a way that the rotational
character of the collective bands is preserved.

In general, it is believed that the collective states of odd nuclei
can be described by using particle-core coupled-type models. The
natural extension of IBM, the Interacting Boson-Fermion Model (IBFM)
\cite{IBFM}, which includes single-particle (fermion) degrees of
freedom in addition to the collective (boson) ones, have provided in
the last decays a unified framework for the description of even-even
and odd-even nuclei distant from closed shell configurations, at
least in the low-angular momentum domain.

For the description of odd$-A$ nuclei, a fermion needs to be coupled
to the $N$ boson system. This can be done by a semimicroscopical
approach which relies on seniority in the nuclear shell model
\cite{IS}. As an alternative to this, in the IBFM approach,
Hamiltonians exhibiting dynamical Bose-Fermi symmetries, that are
analytically solvable \cite{IBFM} are constructed. Thus, the
extension of the IBM for the case of odd mass nuclei leads to the
group structure $U^{B}(6)\otimes U^{F}(m)$ (IBFM-1) or $U_{\pi
}^{B}(6)\otimes U_{\nu }^{B}(6)\otimes U^{F}(m)$ (IBFM-2), where
$m=\sum_{j}(2j+1)$ is the dimension of the single-particle space.
Obviously, in the general case for arbitrary $m-$ values, analytical
expressions for the nuclear levels would be too cumbersome and will
contain too many parameters. Moreover, orbitals higher in energy
than those of the valence shell might play a role and have to be
considered (for example, in the $Sp(6,R)$ model), thus breaking the
symmetric scheme. Therefore, numerical calculations have to be
performed with schematic Hamiltonians. These deficiencies, motivate
the development of the new extension of the IVBM, which will be
based on the success of the boson description of the even-even
nuclei, but will include the fermion degrees of freedom in a simple
and straightforward way, that still leads to exact analytic
solutions.

In the early 1980s, a boson-number-preserving version of the
phenomenological algebraic Interacting Vector Boson Model (IVBM)
\cite{IVBM} was introduced and applied successfully \cite{IVBMrl} to
a description of the low-lying collective rotational spectra of the
even-even medium and heavy mass nuclei. With the aim of extending
these applications to incorporate new experimental data on states
with higher spins and to incorporate new excited bands, we explored
the symplectic extension of the IVBM \cite{GGG}, for which the
dynamical symmetry group is $Sp(12,R).$ This extension is realized
from, and has its physical interpretation over basis states of its
maximal compact subgroup $U\left(6\right)\subset Sp(12,R)$, and
resulted in the description of various excited bands of both
positive and negative parity of complex systems exhibiting
rotation-vibrational spectra \cite{Spa}. With the present work we
extend the earlier applications of IVBM for the description of the
ground and first excited positive and/or negative bands of odd mass
nuclei. In order to do this we propose a new dynamical symmetry
which is applied to real odd nuclear systems.

Thus, it is the purpose of this paper to bring spin explicitly into
the symplectic IBVM. We approach the problem by considering the
simplest physical picture in which a particle (or quasiparticle)
with intrinsic spin taking a single $j-$value $j$ is coupled to an
even-even nucleus whose states belong to an $Sp(12,R)$ irrep.
Nevertheless, the results for the energy spectra and the intraband
transitions between the states of the ground state band obtained in
this simplified version of the model agree rather well with the
experimental data.

\section{The even-even core}

The IVBM is based on the introduction of two kinds of vector bosons
(called $p$ and $n$ bosons), that \textquotedblleft built
up\textquotedblright\ the collective excitations in the nuclear
system. The creation operators of these bosons are assumed to be
$SO(3)$ vectors and they transform according to two independent
fundamental representations (1,0) of the group $SU(3)$ . These
bosons form a \textquotedblright pseudospin\textquotedblright\
doublet of the $U(2)$ group and differ in their \textquotedblleft
pseudospin\textquotedblright\ projection $\alpha =\pm \frac{1}{2}$.
We want to point out that these vector bosons should be considered
as "building blocks" generating appropriate algebraic structures
rather than real correlated fermion pairs coupled to angular
momentum $l=1$.

The algebraic structure of the IVBM is realized in terms of creation
and annihilation operators $u_{m}^{+}(\alpha )$, $u_{m}(\alpha )$
($m=0,\pm 1$). The later are related to the cyclic coordinates
$x_{\pm 1}(\alpha )=\mp \frac{1}{\sqrt{2}}(x_{1}(\alpha )\pm
ix_{2}(\alpha )), x_{0}(\alpha )=x_{3}(\alpha )$, and their
associated momenta $q_{m}(\alpha )=-i\partial /\partial x^{m}(\alpha
)$ in the standard way
\begin{eqnarray}
u_{m}^{+}(\alpha ) &=&\frac{1}{\sqrt{2}}(x_{m}(\alpha
)-iq_{m}(\alpha )),
\label{cro} \\
u_{m}(\alpha ) &=&(u_{m}^{+}(\alpha ))^{\dagger }),  \nonumber
\end{eqnarray}%
where $x_{i}(\alpha )\ i=1,2,3$ are Cartesian coordinates of a
quasi-particle vectors with an additional index - the projection of
the ``pseudo-spin" $\alpha =\pm \frac{1}{2}$. The bilinear products
of the creation and annihilation operators of the two vector bosons
(\ref{cro}) generate the boson representations of the non-compact
symplectic group $ Sp(12,R)$ \cite{IVBM}:
\begin{eqnarray}
F_{M}^{L}(\alpha ,\beta ) &=& {\sum }_{k,m}C_{1k1m}^{LM}u_{k}^{+}(
\alpha )u_{m}^{+}(\beta ),  \nonumber \\
G_{M}^{L}(\alpha ,\beta ) &=&{\sum }_{k,m}C_{1k1m}^{LM}u_{k}(\alpha
)u_{m}(\beta ),  \label{pairgen}
\end{eqnarray}
\begin{equation}
A_{M}^{L}(\alpha, \beta )={\sum }_{k,m}C_{1k1m}^{LM}u_{k}^{+}(\alpha
)u_{m}(\beta ),  \label{numgen}
\end{equation}
where $C_{1k1m}^{LM}$, which are the usual Clebsch-Gordon
coefficients for $L=0,1,2$ and $M=-L,-L+1,...L$, define the
transformation properties of (\ref{pairgen}) and (\ref{numgen})
under rotations. The commutation relations between the pair creation
and annihilation operators \ (\ref{pairgen}) and the number
preserving operators (\ref{numgen}) are given in \cite{IVBM}.

Being a noncompact group, the representations of $Sp(12,R)$ are of
infinite dimension, which makes it impossible to diagonalize the
most general Hamiltonian. When restricted to the group $U^{B}(6)$,
each irrep of the group $Sp^{B}(12,R)$ decomposes into irreps of the
subgroup characterized by the partitions \cite{GGG},\cite{Q1}:
\begin{equation*}
\lbrack N,{0}^5\rbrack_{6}\equiv \lbrack N]_{6},\qquad
\end{equation*}%
where $N=0,2,4,\ldots$ (even irrep) or $N=1,3,5,\ldots$ (odd irrep).
The subspaces $[N]_{6}$ are  finite dimensional, which simplifies
the problem of diagonalization. Therefore the complete spectrum of
the system can be calculated through the diagonalization of the
Hamiltonian in the subspaces of all the unitary irreducible
representations (UIR) of $U(6)$, belonging to a given UIR of
$Sp(12,R)$, which further clarifies its role of a group of dynamical
symmetry. Since $N$ is the number of collective excitations
(phonons) rather than real nucleon pairs, in the present paper we
consider only the even irrep of $Sp(12,R)$.

The most general one and two-body Hamiltonian can be expressed in
terms of symplectic generators. In general, such rather general
Hamiltonian has to be diagonalized numerically to obtain the energy
eigenvalues and wave functions. There exist, however, special
situations in which the eigenvalues can be obtained in closed,
analytical form. These special solutions provide a framework in
which energy spectra and other nuclear properties can be interpreted
in a qualitative way. These situations correspond to dynamical
symmetries of the Hamiltonian.

The Hamiltonian, corresponding to the unitary limit of IVBM
\cite{GGG}
\begin{equation}
Sp(12,R)\supset U(6)\supset U(3)\otimes U(2)\supset O(3)\otimes
(U(1)\otimes U(1)),  \label{uchain}
\end{equation}%
expressed in terms of the first and second order invariant operators
of the different subgroups in the chain (\ref{uchain}) is
\cite{GGG}:
\begin{equation}
H=aN+bN^{2}+\alpha _{3}T^{2}+\beta _{3}L^{2}+\alpha _{1}T_{0}^{2}.
\label{Hrot}
\end{equation}
$H$ (\ref {Hrot}) is obviously diagonal in the basis
\begin{equation}
\mid [N]_{6};(\lambda,\mu);KLM;T_{0}\rangle \equiv \ \mid
(N,T);KLM;T_{0}\rangle, \label{bast}
\end{equation}
labelled by the quantum numbers of the subgroups of the chain
(\ref{uchain}). Its eigenvalues are the energies of the basis states
of the boson representations of $ Sp(12,R)$:
\begin{eqnarray}
E((N,T),L,T_{0}) &=& aN + bN^{2} + \alpha_{3}T(T+1)  \nonumber \\
&+& \beta_{3}L(L+1)+\alpha_{1}T_{0}^{2}.  \label{Erot}
\end{eqnarray}

The non-compact group $Sp(12,R)$ has a Jordan (three grading)
decomposition with respect to its maximal compact subgroup $U(6)$.
Its Lie algebra $g$ can be decomposed as a vector space direct sum:
\[
g=g_{-}\oplus g_{0}\oplus g_{+}.
\]%
Every unitary lowest weight representation of $Sp(12,R)$ can be
constructed by acting consequently on the boson lowest weight state
(LWS) $\mid \Omega \ \rangle _{B}$, transforming in a definite
$U(6)$ representation, with the raising generators $F_{M}^{L}(\alpha
,\beta )$ which belong to the $g_{+}$ space. This action generates
an infinite set of states (\ref{bast}) that form the basis of a
unitary lowest weight representation of $Sp(12,R)$. If the LWS $\mid
\Omega \ \mathbf{\rangle }_{B}$ transforms irreducibly under $U(6)$,
then the corresponding unitary representation of $Sp(12,R)$ is also
irreducible. The unitary lowest weight irreducible representation of
$Sp(12,R)$ can therefore be uniquely labeled by the $U(6)$ labels of
their lowest weight states. In the boson space there are only two
nonequivalent irreducible lowest weight states, namely, the (boson)
vacuum
\begin{equation}
\mid \Omega \ \rangle _{B} =\mid 0\ \mathbf{\rangle }_{B}\
\label{boson vac}
\end{equation}%
and the "one-particle" state
\begin{equation}
\mid \Omega \ \rangle _{B} = u_{k}^{\dagger}(\alpha) \mid 0\
\rangle_{B}\ .  \label{OPBS}
\end{equation}
The construction of the symplectic basis for the even IR of
$Sp(12,R)$, which can be obtained by action of the fully symmetric
coupled powers of raising operators $F_{M}^{L}(\alpha ,\beta )$ on
the on vacuum state (\ref{boson vac}), is given in details in
\cite{GGG}. The $Sp(12,R)$ classification scheme for the $ SU(3)$
boson representations for even value of the number of bosons $N$ \
is shown on Table I in Ref. \cite{GGG} (see also Table
\ref{BasTab}).

\begin{figure}[t]
\centering
\includegraphics[width=12cm,height=8cm]{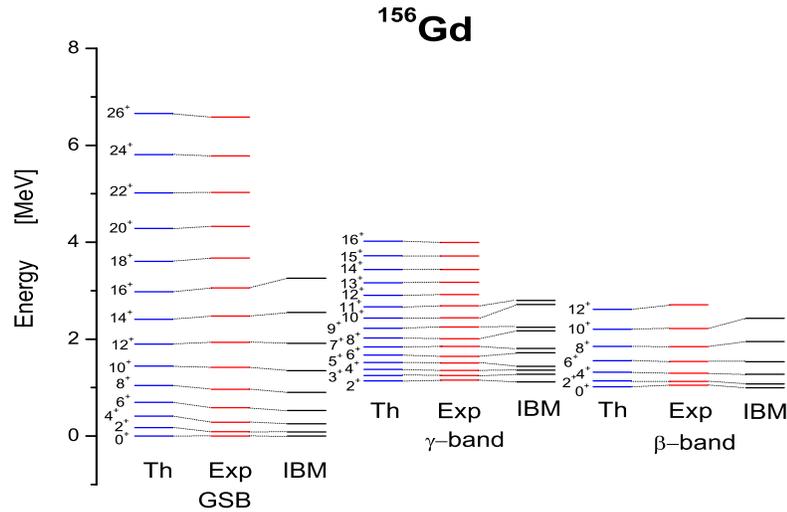}
\caption{Comparison of the theoretical and experimental energies for
the ground and first excited bands of $^{156}$Gd.} \label{Gde}
\end{figure}

\begin{figure}[t]
\centering
\includegraphics[width=12cm,height=8cm]{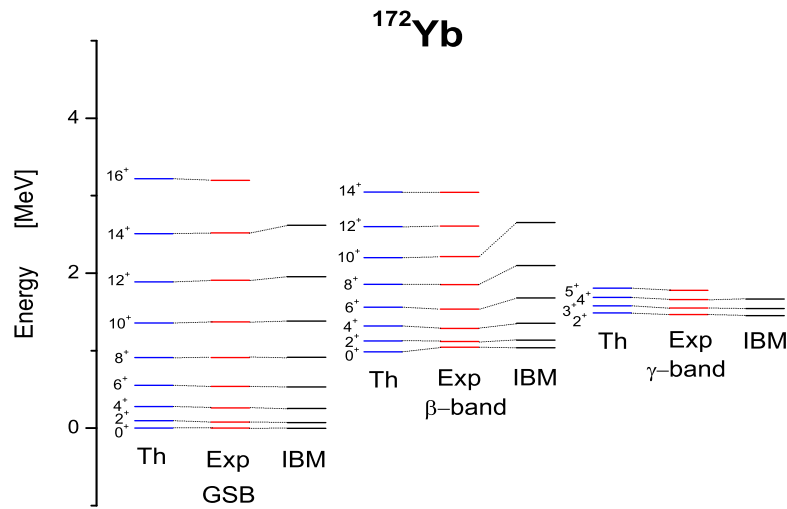}
\caption{The same as Fig. \ref{Gde}, but for $^{172}$Yb.}
\label{Ybe}
\end{figure}

\begin{figure}[t]
\centering
\includegraphics[width=12cm,height=8cm]{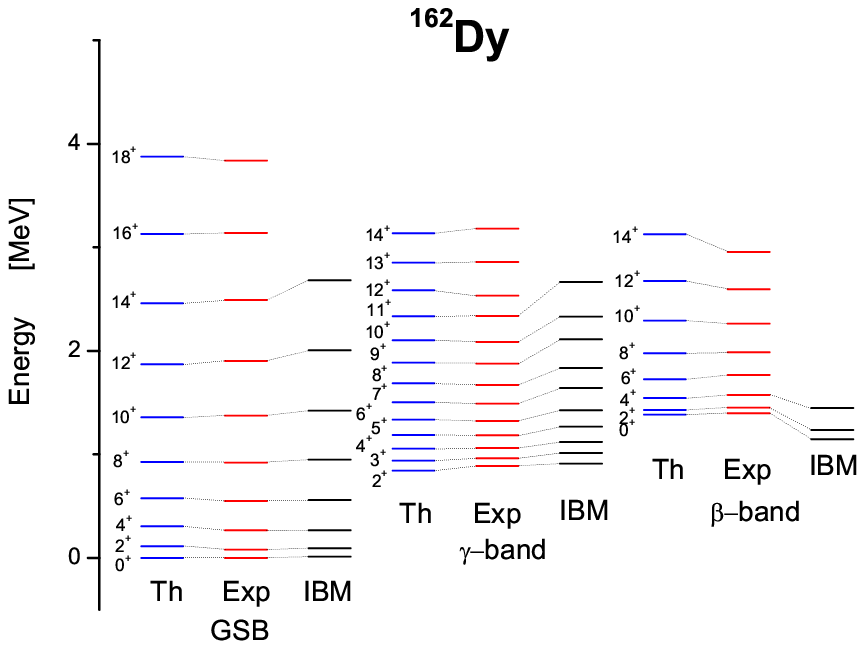}
\caption{The same as Fig. \ref{Gde}, but for $^{162}$Dy.}
\label{Dye}
\end{figure}

The most important application of the $U^{B}(6)\subset $
$Sp^{B}(12,R)$ limit of the theory is the possibility it affords for
describing both even and odd parity bands up to very high angular
momentum \cite{GGG}. In order to do this we first have to identify
the experimentally observed bands with the sequences of basis states
of the even $Sp(12,R)$ irrep (Table \ref{BasTab}). As we deal with
the symplectic extension of the boson representations of the number
preserving $U^{B}(6)$ symmetry we are able to consider all even
eigenvalues of the number of vector bosons $N$ with the
corresponding set of pseudospins $T$, which uniquely define the
$SU^{B}(3)$ irreps $(\lambda, \mu)$. The multiplicity index $K$
appearing in the final reduction to the $SO(3)$ is related to the
projection of $L$ on the body fixed frame and is used with the
parity ($\pi $)\ to label the different bands ($K^{\pi } $) in the
energy spectra of the nuclei. For the even-even nuclei we have defined
the parity of the states as $\pi_{core} =(-1)^{T}$ \cite{GGG}. This
allowed us to describe both positive and negative bands.

Further, we use the algebraic concept of \textquotedblleft
yrast\textquotedblright\ states, introduced in \cite{GGG}. According
to this concept we consider as yrast states the states with given
$L$, which minimize the energy (\ref{Erot}) with respect to the
number of vector bosons $N$ that build them. Thus the states of the
ground state band (GSB) were identified with the $SU(3)$ multiplets
$(0,\mu )$ \cite{GGG}. In terms of $(N,T)$ this choice corresponds
to $(N=2\mu ,T=0)$ and the sequence of states with different numbers
of bosons $N=0,4,8,\ldots $ and pseudospin $T=0$ (and also
$T_{0}=0$). Hence the minimum values of the energies (\ref{Erot})
are obtained at $N=2L$.

The presented mapping of the experimental states onto the $SU(3)$
basis states, using the algebraic notion of yrast states, is a
particular case of the so called "stretched" states \cite{Stretch}.
The latter are defined as the states with ($\lambda_{0}+2k,\mu_{0}$)
or ($\lambda_{0},\mu_{0}+ k$), where $N_{i}=\lambda_{0}+2\mu_{0}$
and $k=0,1,2,3, \ldots$.

It was established \cite{GGD} that the correct placement of the
bands in the spectrum strongly depends on their bandheads'
configuration, and in particular, on the minimal or initial number
of bosons, $N = N_{i}$, from which they are built. The latter
determines the starting position of each excited band.

Thus, for the description of the different excited bands, we first
determine the $N_{i}$ of the band head structure and develop the
corresponding excited band over the stretched $SU(3)$ multiplets.
This corresponds to the sequence of basis states with
$N=N_{i},N_{i}+4,N_{i}+8,\ldots$ ($\Delta N=4$). The values of $T$
for the first type of stretched states ($\lambda-$changed) are
changed by step $\Delta T=2$, whereas for the second type
($\mu-$changed) $-T$ is fixed so that in both cases the parity is
preserved even or odd, respectively. For all presented even-even nuclei,
the states of the corresponding $\beta -$ and $\gamma -$ bands are
associated with the stretched states of the first type ($\lambda-$
changed).

The odd-A nuclei $^{157}Gd$, $^{173}Yb$ and $^{163}Dy$, to which we
apply our model, can be considered as a particle coupled to the
even-even cores $^{156}Gd$, $^{172}Yb$ and $^{162}Dy$, respectively.
We determine the values of the five phenomenological model
parameters $a, b, \alpha_{3}, \beta_{3}, \alpha_{1}$ by fitting the
energies of the ground and few excited bands ($\gamma-$ and/or
$\beta-$ bands) of the even-even nuclei to the experimental data
\cite{exp}, using a $\chi^{2}$ procedure. The theoretical
predictions for the even core nuclei are presented in the Figures
\ref{Gde}$-$\ref{Dye}. For comparison, the predictions of IBM (with
$4$ adjustable parameters) are also shown. The IBM results for
$^{156}Gd$ and $^{162}Dy$,$^{172}Yb$ are extracted from Refs.
\cite{IBFMGd} and \cite{IBFMYb}, respectively. From the figures one
can see that the calculated energy levels agree rather well up to
very high angular momenta with the observed data. One can see also
that for high spins ($L\geq 10-14$), where the deviations of the IBM
predictions become more significant, the structure of the energy
levels of the GSB ($\beta-$ and $\gamma-$bands) is reproduced rather
well.

\section{The inclusion of spin}

Underlying the conventional nuclear shell model is the idea that the
low-lying states of nuclei can be restricted to a valence space of
states obtained by putting nucleons into a finite set of
single-particle states indexed $i=1,\ldots,\Omega$;  i. e. the $M$
valence-particle Hilbert space is the anti-symmetrized (exterior)
product of $M$ copies of an $\Omega-$dimensional single-nucleon
Hilbert space. This space carries a sum of two irreducible
representations of the fermion pair algebra $SO(2\Omega)$
\cite{pair}. The set of all even fermion states span an irreducible
representation of the $SO(2\Omega)$ algebra and the set of all
states of odd fermion number span another irreducible
representation.

Thus, in order to incorporate the intrinsic spin degrees of freedom
into the symplectic IVBM, we extend the dynamical algebra of
$Sp(12,R)$ to the orthosymplectic algebra of $OSp(2\Omega/12,R)$.
For this purpose we introduce a particle (quasiparticle) with spin
$j$ and consider a simple core plus particle picture. Thus, in
addition to the boson collective degrees of freedom (described by
dynamical symmetry group $Sp(12,R)$) we introduce creation and
annihilation operators $a_{m}^{\dag}$ and $a_{m}$ ($m=-j,\ldots,j$),
which satisfy the anticommutation relations
\begin{eqnarray}
\{a_{m}^{\dag},a_{m'}^{\dag}\} &=&\{a_{m},a_{m'}\}=0,  \nonumber \\
\{a_{m},a_{m'}^{\dag}\} &=&\delta _{mm'}. \label{anticom}
\end{eqnarray}%
All bilinear combinations of \ $a_{m}^{+}$ and $a_{m'}$, namely
\begin{eqnarray}
f_{mm'} &=&a_{m}^{\dag}a_{m'}^{\dag}, \ \ m\neq m' \nonumber \\
g_{mm'} &=&a_{m}a_{m'}, \ \ m\neq m'; \label{O4 gen} \\
C_{mm'} &=&(a_{m}^{\dag}a_{m'}-a_{m'}a_{m}^{\dag})/2  \label{U2 gen}
\end{eqnarray}%
generate the (Lie) fermion pair algebra of $SO^{F}(2\Omega)$. Their
commutation relations are:
\begin{eqnarray*}
\lbrack g_{mn},C_{m^{\prime }n^{\prime }}] &=&\delta _{nm^{\prime
}}g_{mn^{\prime }}-\delta _{mm^{\prime }}g_{nn^{\prime }}, \\
\lbrack C_{mn},f_{n^{\prime }n^{\prime }}] &=&\delta _{nm^{\prime
}}g_{mn^{\prime }}-\delta _{nm^{\prime }}g_{mn^{\prime }}, \\
\lbrack g_{mn},f_{m^{\prime }n^{\prime }}] &=&-\delta _{mm^{\prime
}}C_{n^{\prime }n}-\delta _{nn^{\prime }}C_{n^{\prime }m}+\delta
_{n^{\prime }m}C_{n^{\prime }n}+\delta _{m^{\prime }n}C_{n^{\prime
}m},
\end{eqnarray*}%
The number preserving operators (\ref{U2 gen}) generate maximal
compact subalgebra of $SO^{F}(2\Omega)$, i.e. $U^{F}(\Omega)$. The
upper (lower) script $B$ or $F$ denotes the boson or fermion degrees
of freedom, respectively.

Making use of the embedding $SU^{F}(2)\subset SO^{F}(2\Omega)$, we
make orthosymplectic (supersymmetric) extension of the IVBM which is
defined through the chain:
\begin{equation}
\begin{tabular}{lllll}
$OSp(2\Omega/12,R)$ & $\supset $ & $SO^{F}(2\Omega)$ & $\otimes $ & $Sp^{B}(12,R)$ \\
&  &  &  & $\ \ \ \ \ \ \Downarrow $ \\
&  & $\ \ \ \ \ \Downarrow $ & $\otimes $ & \ $\ U^{B}(6)$ \\
&  &  &  & $\ \ \ \ \ N$ \\
&  &  &  & $\ \ \ \ \ \ \Downarrow $ \\
&  & $SU^{F}(2)$ & $\otimes $ & $SU^{B}(3)\otimes U_{T}^{B}(2)$ \\
&  & $\ \ \ \ \ j$ &  & $(\lambda ,\mu )\Longleftrightarrow (N,T)~$ \\
&  & \multicolumn{1}{r}{$\searrow $} &  & $\ \ \ \ \ \ \Downarrow $ \\
&  &  & $\otimes $ & $SO^{B}(3)\otimes U(1)$ \\
&  &  &  & $~~~L\qquad \qquad T_{0}$ \\
&  &  & $\Downarrow $ &  \\
&  & $Spin^{BF}(3)$ & $\supset $ & $Spin^{BF}(2),$ \\
&  & $~~~~~J$ &  & $~~~~~J_{0}$%
\end{tabular}
\label{chain}
\end{equation}%
where bellow the different subgroups the quantum numbers
characterizing their irreducible representations are given. Here
with $Spin^{BF}(n)$ ($n=2,3$) is denoted the universal covering group of
the $SO(n)$. From (\ref{chain}) it can be seen that the coupling of
the boson and fermion degrees of freedom is done on the level of the
angular momenta. We want to stress, however, that although the
formal "coupling" is done at the "final" stage, the present
situation is not identical to that of IBFM. In the latter the
even-even core, to which an odd unpaired nucleon is coupled to, is
considered as "inert". In the present approach since the
(ortho)symplectic structures are taken into account (allowing for
the change of number of phonon excitations $N$), the core is not
anymore inert. Physically, this does not correspond to the weak
coupling limit (as should be if $N$ was fixed) between the core and
particle as it is in the case of IBFM (on this level of coupling).

\section{Application of the new dynamical symmetry}
\subsection{The energy spectrum}

In this paper we expand the earlier application of the IVBM
\cite{GGG}, developed for the description of the collective bands of
even-even nuclei, in order to include in our considerations the case
of odd mass nuclei.

We can label the basis states according to the chain (\ref{chain})
as:
\begin{equation}
|~[N]_{6};(\lambda,\mu);KL;j;JJ_{0};T_{0}~\rangle \equiv
|~[N]_{6};(N,T);KL;j;JJ_{0};T_{0}~\rangle ,  \label{Basis}
\end{equation}
where $[N]_{6}-$ is the $U(6)$ labeling quantum number,
$(\lambda,\mu)-$ are the $ SU(3)$ quantum numbers characterizing the
core excitations, $K$ is the multiplicity index in the reduction
$SU(3)\subset SO(3)$, $L$ is the core angular momentum, $j-$the spin
of the odd particle, $J,J_{0}$ are the total (coupled) angular
momentum and its third projection, and $T$,$T_{0}$ are the
pseudospin and its third projection, respectively. Since the
$SO(2\Omega)$ label is irrelevant for our application, we drop it in
the states (\ref{Basis}).

The Hamiltonian can be written as linear combination of the Casimir
operators of the different subgroups in (\ref{chain}): \
\begin{equation}
H = aN+bN^{2}+\alpha _{3}T^{2}+\beta _{3}^{\prime
}L^{2}+\alpha_{1}T_{0}^{2} + \eta j^{2}+\gamma ^{\prime }J^{2}+\zeta
J_{0}^{2} \label{Hamiltonian}
\end{equation}
and it is obviously diagonal in the basis (\ref{Basis}) labeled by
the quantum numbers of their representations. Then the eigenvalues
of the Hamiltonian (\ref{Hamiltonian}), that yield the spectrum of
the odd mass system are:

\begin{eqnarray}
E(N;T,T_{0};L,j;J,J_{0})=&aN+bN^{2}+\alpha _{3}T(T+1)+
\beta _{3}^{\prime }L(L+1)+\alpha _{1}T_{0}^{2} \nonumber \\
&+\eta j(j+1)+\gamma ^{\prime }J(J+1)+\zeta J_{0}^{2}.\label{Energy}
\end{eqnarray}

We note that only the last three terms of (\ref{Hamiltonian}) come
from the orthosymplectic extension. But since only one fermion
($M=1$) is considered (and $j$ is fixed), the $j-$term in
(\ref{Energy}) is just additive constant and can be dropped. (The
presence of the latter should only rescale the values of the rest
model parameters.) Thus, for the description of the excitation
spectra of odd-mass nuclei only two new parameters are involved in
the fitting procedure. We choose parameters $\beta _{3} ^{\prime
}=\frac{1}{2}\beta _{3}$ and $\gamma ^{\prime }=\frac{1}{2}\gamma$
instead of $\beta _{3}$ and $\gamma$ in order to obtain the
Hamiltonian form of ref. \cite{GGG} (setting $\beta _{3} =\gamma$),
when for the case $j=0$ (hence $J=L$) we recover the symplectic
structure of the IVBM.

The infinite set of basis states classified according to the
reduction chain (\ref{chain}) are schematically shown in Table
\ref{BasTab}. The fourth and fifth columns show the $SO^{B}(3)$
content of the $SU^{B}(3)$ group, given by the standard Elliott's
reduction rules \cite{Elliott}, while in the next column are given
the possible values of the common angular momentum $J$, obtained by
coupling of the orbital momentum $L$ with the spin $j$. The latter
is vector coupling and hence all possible values of the total angular
momentum $J$ should be considered. For simplicity, only the maximally
aligned ($J=L+j$) and maximally antialigned ($J=L-j$) states are
illustrated in Table \ref{BasTab}.

\begin{center}
\begin{table}[h]
\caption{Classification scheme of basis states (\protect\ref{Basis})
according the decompositions given by the chain
(\protect\ref{chain}).}
\smallskip \centering{\footnotesize \renewcommand{\arraystretch}{1.25}
\begin{tabular}{l|l|l|l|l|l}
\hline\hline $N$ & $T$ & $(\lambda ,\mu )$ & $K$ & $L$ & \quad \quad
\quad \quad \quad \quad \quad $J=L\pm j$ \\ \hline\hline $0$ & $0$ &
$(0,0)$ & $0$ & $0$ & $j$ \\ \hline\hline $2$ & $1$ & $(2,0)$ & $0$
& $0,2$ & $j;\ 2\pm j$ \\ \cline{2-6} & $0$ & $(0,1)$ & $0$ & $1$ &
$1\pm j$ \\ \hline\hline & $2$ & $(4,0)$ & $0$ & $0,2,4$ & $j;\ 2\pm
j;4\pm j$ \\ \cline{2-6}
$4$ & $1$ & $(2,1)$ & $1$ & $1,2,3$ & $1\pm j;\ 2\pm j;\ 3\pm j$ \\
\cline{2-6} & $0$ & $(0,2)$ & $0$ & $0,2$ & $ j;\ 2\pm j$ \\
\hline\hline
& $3$ & $(6,0)$ & $0$ & $0,2,4,6$ & $ j;\ 2\pm j;\ 4\pm j;6\pm j$ \\
\cline{2-6} & $2$ & $(4,1)$ & $1$ & $1,2,3,4,5$ &
\begin{tabular}{l}
$1\pm j;\ 2\pm j;\ 3\pm j;$ \\
$4\pm j;\ 5\pm j$%
\end{tabular}
\\ \cline{2-6}
$6$ & $1$ & $(2,2)$ & $2$ & $2,3,4$ & $2\pm j;\ 3\pm j;\ 4\pm j$ \\
\cline{2-6} &  &  & $0$ & $0,2$ & $ j;\ 2\pm j$ \\ \cline{2-6} & $0$
& $(0,3)$ & $0$ & $1,3$ & $1\pm j;\ 3\pm j$ \\ \hline\hline & $4$ &
$(8,0)$ & $0$ & $0,2,4,6,8$ &
\begin{tabular}{l}
$ j;\ 2\pm j;\ 4\pm j;$ \\
$6\pm j;\ 8\pm j$%
\end{tabular}
\\ \cline{2-6}
& $3$ & $(6,1)$ & $1$ & $1,2,3,4,5,6,7$ &
\begin{tabular}{l}
$1\pm j;\ 2\pm j;\ 3\pm j;$ \\
$4\pm j;\ 5\pm j;6\pm j;\ $ \\
$7\pm j;8\pm j$%
\end{tabular}
\\ \cline{2-6}
& $2$ & $(4,2)$ & $2$ & $2,3,4,5,6$ &
\begin{tabular}{l}
$2\pm j;\ 3\pm j;\ 4\pm j;$ \\
$5\pm j;\ 6\pm j$%
\end{tabular}
\\ \cline{2-6}
$8$ &  &  & $0$ & $0,2,4$ & $ j;\ 2\pm j;\ 4\pm j$ \\ \cline{2-6}
& $1$ & $(2,3)$ & $2$ & $2,3,4,5$ & $2\pm j;\ 3\pm j;\ 4\pm j;5\pm j$ \\
\cline{2-6} &  &  & $0$ & $1,3$ & $1\pm j;\ 3\pm j$ \\ \cline{2-6} &
$0$ & $(0,4)$ & $0$ & $0,2,4$ & $ j;2\pm j;\ 4\pm j$ \\
\hline\hline
$\vdots $ & $\vdots $ & $\vdots $ & $\vdots $ & $\vdots $ & $\vdots $%
\end{tabular}
} \label{BasTab}
\end{table}
\end{center}

The basis states (\ref{Basis}) can be considered as a result of the
coupling of the orbital $\mid (N,T);KLM;T_{0}\rangle$ (\ref{bast})
and spin $\phi_{jm}$ wave functions. Then, if the parity of the
single particle is $\pi_{sp}$, the parity of the collective states
of the odd$-A$ nuclei will be $\pi=\pi_{core} \pi_{sp}$. Thus, the
description of the positive and/or negative parity bands requires
only the proper choice of the core band heads, on which the
corresponding single particle is coupled to, generating in this way
the different odd$-A$ collective bands. Our choice is based on the
fact, which has been always understood in nuclear physics, that well
defined rotational bands can exist only when they are adiabatic
relative to other degrees of freedom. In this way (in adiabatic
approximation) the single particle is dragged around in the core
field (which corresponds to the "strong" coupling limit as is in our case)
and the combined system is essentially a new rotor with
slightly different bulk properties, such as moment of inertia, etc.

Further in the present considerations, the yrast conditions yield
relations between the number of bosons $N$ and the coupled angular
momentum $J$ that characterizes each collective state. For example,
the collective states of the GSB $K_{J} ^{\pi}=\frac{3}{2}^{-}$ are
identified with the $SU(3)$ multiplets $(0,\mu )$  which yield the
sequence $N=2(J-j)=0,2,4,\ldots$ for the corresponding values
$J=\frac{3}{2},\frac{5}{2},\frac{7}{2},...$. The pseudospin for
the $SU(3)$ multiplets $(0,\mu )$ is $T=0$ and hence
$\pi_{core}=(-1)^{T}=(+)$. Here it is assumed that
the single particle has $j=3/2$ and parity $\pi_{sp}=(-)$, so
that the common parity $\pi$ is also negative.

For the description of the different excited bands, we first
determine the $N_{i}$  of the band head structure and then we map
the states of the corresponding band onto the sequence of basis
states with $N=N_{i},N_{i}+2,N_{i}+4,\ldots$ ($\Delta N=2$) and
$T=even=fixed$ or $T=odd=fixed$, respectively. This choice
corresponds to the stretched states of the second type
($\mu-$changed).

We will point out that the (ortho)symplectic structure of the model
space gives us rather rich possibilities to map experimentally
observed states onto the basis states. Thus, another possibility of
developing the sequence of band's states is to take again
$N=N_{i},N_{i}+4,N_{i}+8,\ldots$ ($\Delta N=4$) but to change
$T=T_{i},T_{i}+2,T_{i}+4,...$ ($\Delta T=2$) in such a way, that the
parity is preserved even or odd, respectively. Such correspondence
takes place for the first type of the stretched states
($\lambda-$changed). In the present application, all the collective
bands under consideration are associated with the stretched states
of second type ($\mu-$changed).

\begin{figure}[t]
\centering
\includegraphics[width=12cm,height=8cm]{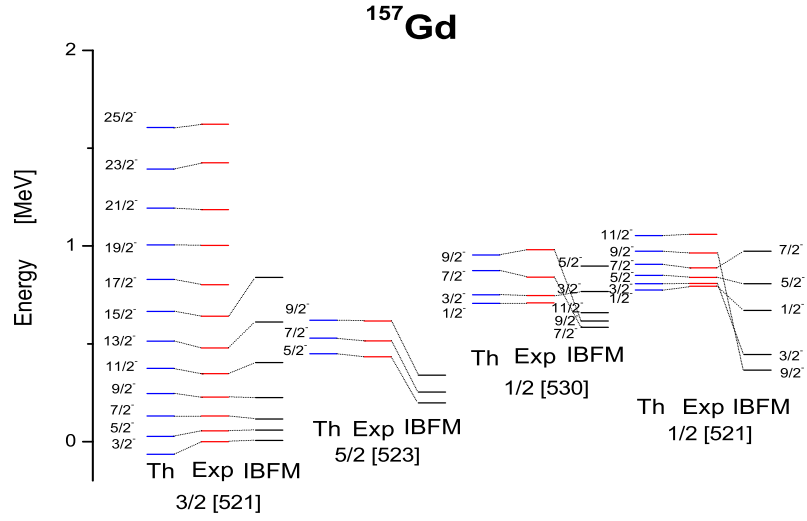}
\caption{Comparison of the theoretical and experimental energies for
the ground and first excited negative parity bands of
$^{157}$Gd.}\label{Gdo}
\end{figure}

\begin{figure}[t]
\centering
\includegraphics[width=12cm,height=8cm]{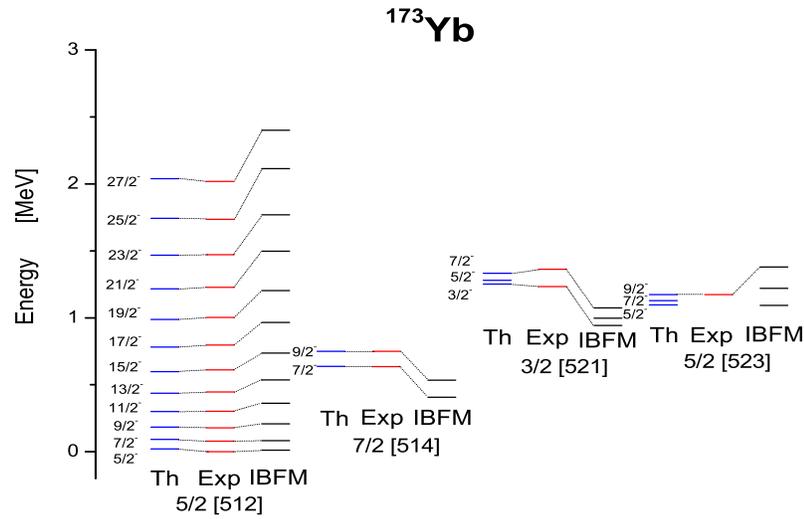}
\caption{The same as Fig. \ref{Gdo}, but for $^{173}$Yb.}\label{Ybo}
\end{figure}

\begin{figure}[t]
\centering
\includegraphics[width=12cm,height=8cm]{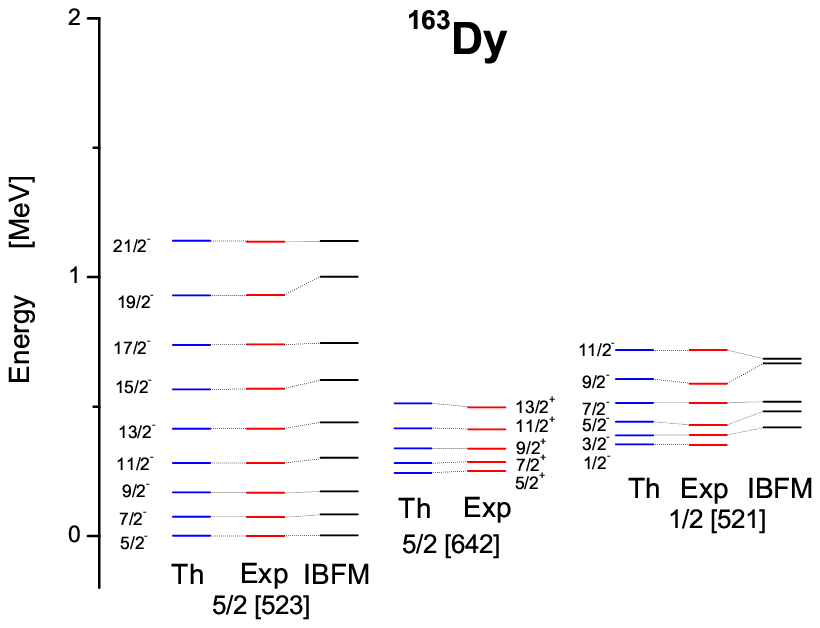}
\caption{Comparison of the theoretical and experimental energies for
the ground and first excited positive and negative parity bands of
$^{163}$Dy.}\label{Dyo}
\end{figure}

The number of adjustable parameters needed for the complete
description of the collective spectra of the odd-A nuclei  is two,
namely $\gamma$ and $\zeta $. They are evaluated by a fit to the
experimental data \cite{exp} of the GSB of the corresponding odd-A
nucleus. The comparison between the experimental spectra for the GSB
and first few excited bands and our calculations using the values of
the model parameters given in Table \ref{TablePar} for the nuclei
$^{157}Gd$, $^{173}Yb$ and $^{163}Dy$ is illustrated in Figures
\ref{Gdo}$-$\ref{Dyo}. The last single particle, which for all of
these rare earth nuclei is a neutron, occupies the major shell
$N=82-126$, where the relevant single particle levels are
$2f_{\frac{7}{2}},2f_{\frac{5}{2}},3p_{\frac{3}{2}},3p_{\frac{1}{2}}$
having odd parity ($\pi_{sp}=-$) (excluding the intruder from the
upper shell with opposite parity). In our considerations we take
into account only the first available single particle orbit $j$
(generating the group $SO(2\Omega)$ with $\Omega=(2j+1)$), which for
the first nucleus implies $j=\frac{3}{2}$, while for the other two
$-$ $j=\frac{5}{2}$. The Nilsson asymptotic quantum numbers
$\Omega[Nn_{3}\Lambda]$ are written bellow each band. One can see
from the figures that the calculated energy levels agree rather well
in general with the experimental data up to very high angular
momenta. For comparison, in the Figures \ref{Gdo}$-$\ref{Dyo} the
IBFM results (obtained by total $7$ adjustable parameters) are also
shown. They are extracted from Refs. \cite{IBFMGd} and
\cite{IBFMYb}, respectively. Note that all calculated levels, for
the bands considered, are in correct order in contrast to IBFM
results (for $^{157}Gd$). Another difference between the IVBM and
IBFM predictions is that in the former the correct placement of all
the band heads is reproduced quite well.

In the Table \ref{TablePar}, the values of $N_{i}$, $T$, $T_{0}$,
$J$, $J_{0}$ and $\chi^{2}$ for each band under consideration are
also given.

\begin{table}[tb]
\caption{{Values of the model parameters.}}%
\centering%
\begin{tabular}{||l|l|l|l|l|l|l|l|l||}
\hline\hline
$Nucl.$ & $bands$ & $N_{i}$ & $T$ & $T_{0}$ & $J$ & $J_{0}$ & $\chi ^{2}$ & $%
parameters$ \\ \hline\hline
$^{157}Gd$ & $GSB:\allowbreak K^{\pi }=3/2^{-}$ & $0$ & $0$ & $0$ & $L+j$ & $%
3/2$ & $0.00100$ & $%
\begin{tabular}{l}
$a=0.03225$ \\
$b=-0.00075$ \\
$\alpha _{3}=0.00332$%
\end{tabular}%
$ \\ \hline $j=3/2$ & $K^{\pi }=5/2^{-}$ & $20$ & $8$ & $4$ & $L-j$
& $5/2$ & $0.00014$ &
\begin{tabular}{l}
$\beta _{3}=0.00998$ \\
$\alpha _{1}=-0.00303$%
\end{tabular}
\\ \hline
& $K^{\pi }=1/2^{-}[530]$ & $20$ & $10$ & $3$ & $L-j$ & $1/2$ & $0.00061$ & $%
\gamma =0.00806$ \\ \hline
& $K^{\pi }=1/2^{-}[521]$ & $24$ & $12$ & $6$ & $L-j$ & $1/2$ & $0.00018$ & $%
\zeta =-0.03549$ \\ \hline\hline
$^{173}Yb$ & $GSB:\allowbreak K^{\pi }=5/2^{-}$ & $0$ & $0$ & $0$ & $L+j$ & $%
5/2$ & $0.00018$ &
\begin{tabular}{l}
$a=0.00716$ \\
$b=-0.00027$ \\
$\alpha _{3}=0.00153$%
\end{tabular}
\\ \hline
$j=5/2$ & $K^{\pi }=7/2^{-}$ & $40$ & $20$ & $5$ & $L-j$ & $7/2$ &
$0.000002$ &
\begin{tabular}{l}
$\beta _{3}=0.01198$ \\
$\alpha _{1}=-0.00605$%
\end{tabular}
\\ \hline
& $K^{\pi }=3/2^{-}$ & $72$ & $36$ & $3$ & $L-j$ & $3/2$ & $0.00034$ & $%
\gamma =0.01281$ \\ \hline
& $K^{\pi }=5/2^{-}[523]$ & $60$ & $30$ & $0$ & $L-j$ & $J$ & $0.000007$ & $%
\zeta =-0.00555$ \\ \hline\hline
$^{163}Dy$ & $GSB:\allowbreak K^{\pi }=5/2^{-}$ & $0$ & $0$ & $0$ & $L+j$ & $%
5/2$ & $0.000004$ &
\begin{tabular}{l}
$a=0.01242$ \\
$b=0.00041$ \\
$\alpha _{3}=0.00170$%
\end{tabular}
\\ \hline
$j=5/2$ & $\allowbreak K^{\pi }=5/2^{+}$ & $22$ & $11$ & $3$ & $L+j$
& $5/2$ & $0.00008$ &
\begin{tabular}{l}
$\beta _{3}=0.01159$ \\
$\alpha _{1}=-0.00658$%
\end{tabular}
\\ \hline
& $\allowbreak K^{\pi }=1/2^{-}$ & $32$ & $16$ & $5$ & $L-j$ & $1/2$ & $%
0.00007$ &
\begin{tabular}{l}
$\gamma =0.01124$ \\
$\zeta =-0.00779$%
\end{tabular}
\\ \hline\hline
\end{tabular}%
\label{TablePar}
\end{table}

\subsection{Electromagnetic transition probabilities}

A successful nuclear model must yield a good description not only of
the energy spectrum of the nucleus but also of its electromagnetic
properties. Calculation of the latter is a good test of the nuclear
model functions. The most important electromagnetic features are the
$E2$ transitions. In this subsection we discuss the calculation of
the $E2$ transition strengths and compare the results with the
available  experimental data.

As was mentioned, in the symplectic extension of the IVBM the
complete spectrum of the system is obtained in all the even
subspaces with fixed $N$- even of the UIR $[N]_{6}$ of $U(6)$,
belonging to a given even UIR of $Sp(12,R)$. The classification
scheme of the $SU(3)$ boson representations for even values of the
number of bosons $N$ was presented in Table \ref{BasTab}.

In this paper we give as an example the evaluation of the $E2$
transition probabilities between the states of the ground state bands $K_{J}
^{\pi}=\frac{3}{2}^{-}$ and $K_{J} ^{\pi}=\frac{5}{2}^{-}$. For both
cases, the states of the GSB are identified with the $SU(3)$
multiplets $(0,\mu )$ and $\mu=L$. This yields the sequence
$N=2(J-j)=0,2,4,\ldots $ for the corresponding values
$J=\frac{3}{2},\frac{5}{2},\frac{7}{2},...$. In terms of $(N,T)$
this corresponds to $(N=2\mu ,T=0)$.

Using the tensorial properties of the $Sp(12,R)$ generators with
respect to (\ref{uchain}) it is
easy to define the $E2$ transition operator  \cite{TP} between the
states of the considered band as:
\begin{equation}
T^{E2}=e\left[A_{[210]_{3}[0]_{2}\quad 00}^{ \lbrack 1-1]_{6}\quad
\quad 20}+\theta ([F\times F]_{(0,2)[0]_{2}\quad 00}^{\quad \lbrack
4]_{6}\quad \, \ 20}+[G\times G]_{(2,0)[0]_{2}\quad 00}^{\quad
\lbrack -4]_{6}\quad \,20}\right].  \label{te2}
\end{equation}%
The first part of (\ref{te2}) is a
$SU(3)$ generator and actually changes only the angular momentum
with $\Delta L=2$.

The tensor product
\begin{eqnarray} \label{FF}
\lbrack F\times F]_{(0,2)[0]_{2}\quad \, 00}^{\quad \lbrack
4]_{6}\quad \quad 20} &=&\sum C_{(2,0)[2]_{2} \, (2,0)[2]_{2}\quad
(0,2)[0]_{2}}^{\quad [2]_{6}\quad \quad \lbrack 2]_{6}\quad \quad
\lbrack
4]_{6}}C_{\,(2)_{3}\quad (2)_{3}\quad (2)_{3}}^{(2,0)\,\,(2,0)\quad(0,2)} \nonumber \\
&& \label{FF} \\
&&\times C_{20\ 20}^{20}C_{1 1\ 1 -1}^{10}\ F_{(2,0)[2]_{2}\ \
11}^{\quad \lbrack 2]_{6}\quad \ 20}F_{(2,0)[2]_{2}\ \ 1-1}^{\quad
\lbrack 2]_{6}\quad \ 20} \nonumber
\end{eqnarray}%
of the operators (\ref{pairgen}) that are the pair raising
$Sp(12,R)$ generators changes the number of bosons by $\Delta N=4$
and $\Delta L=2$. It is obvious that this term in $T^{E2}$
(\ref{te2}) comes from the symplectic extension of the model. In
(\ref{te2}) $e$ is the effective boson charge.

The transition probabilities are by definition $SO(3)$ reduced
matrix elements of transition operators $T^{E2}$ (\ref{te2}) between
the $|i\rangle -$initial and $|f\rangle -$final collective states
(\ref{Basis})
\begin{equation}
B(E2;J_{i}\rightarrow J_{f})=\frac{1}{2J_{i}+1}\mid \langle \quad
f\parallel T^{E2}\parallel i\quad \rangle \mid ^{2}. \label{deftrpr}
\end{equation}
As was mentioned, the basis states (\ref{Basis}) can be considered
as a result of the coupling of the orbital $\mid
(N,T);KLM;T_{0}\rangle$ (\ref{bast}) and spin $\phi_{jm}$ wave
functions. Since the spin $j$ is simply vector coupled to the
orbital momentum $L$, the action of the transition operator $T^{E2}$
concerns only the orbital part of the basis functions (\ref{Basis}).

In Ref. \cite{TP} it is shown that the two main  types of $B(E2)$
behavior -- the enhancement or the reduction of the $B(E2)$ values
within the GSB $K^{\pi}=0^{+}$, can be reproduced simply by the
change of the sign of $\theta$. The strongly enhanced values which
are an indication for increased collectivity in the high angular
momentum domain are easily obtained for positive values of the
parameter $\theta$. For negative values of the parameter $\theta$ we
obtain behavior similar to that of the standard $SU(3)$ one and it
can be used to reproduce the well known cutoff effect. Such
saturation effect is also characteristic feature of the IBM based
calculations in its $SU(3)$ limit. It is shown also that although
the coefficient in front of symplectic term is some orders of
magnitude smaller than the $SU(3)$ contribution to the transition
operator its role in reproducing the correct behavior (with or
without cutoff) of the transition probabilities between the states
of the GSB band is very important. For more details concerning
discussed behavior of the $B(E2)$ values see \cite{TP}.

\begin{figure}[t]
\centering
\includegraphics[width=12cm,height=10cm]{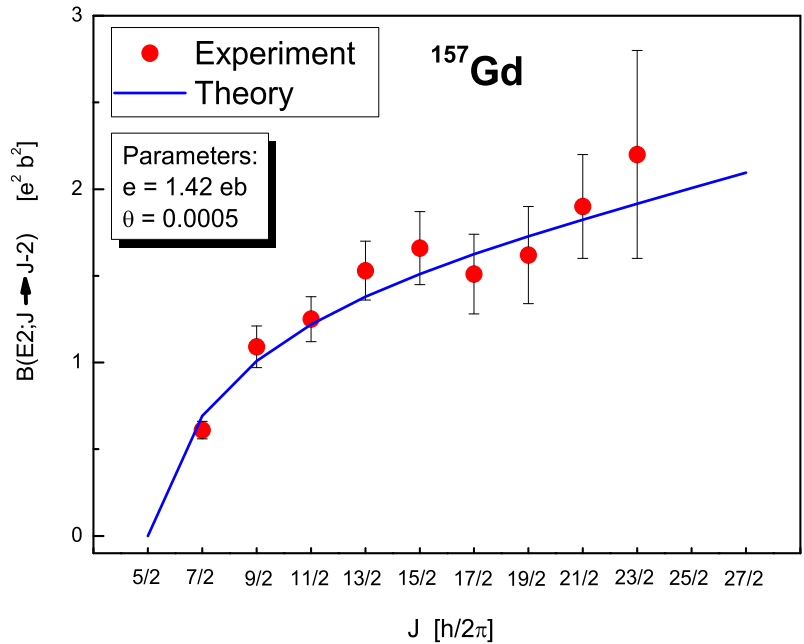}
\caption{(Color online) Comparison of the theoretical and
experimental values for the $B(E2)$ transition probabilities for the
$^{157}Gd$.} \label{Gdt}
\end{figure}

\begin{figure}[t]
\centering
\includegraphics[width=12cm,height=10cm]{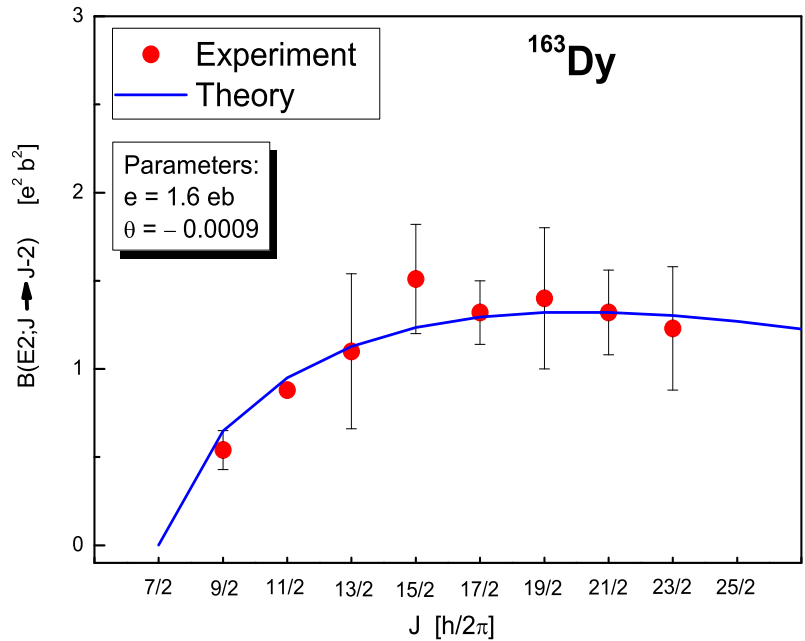}
\caption{(Color online) The same as Fig. \ref{Gdt}, but for
$^{163}$Dy.} \label{Dyt}
\end{figure}

In order to prove the correct predictions following from our
theoretical results we apply the theory to real nuclei for which
there is available experimental data for the transition
probabilities \cite{ExpTran} between the states of the ground bands
up to very high angular momenta. The application actually consists
of fitting the two parameters of the transition operator $T^{E2}$
(\ref{te2}) to the experiment for each of the considered bands. The
$B(E2)$ strengths between the negative parity states of the GSB, as
were attributed to the $SU(3)$ symmetry-adapted basis states of the
model, are calculated. For this $SU(3)$ multiplets, the procedure
for their calculations actually coincides with that given in
\cite{TP}. The theoretical predictions for the nuclei $^{157}Gd$ and
$^{163}Dy$ are compared with the experimental data in Figures
\ref{Gdt} and \ref{Dyt}. From the figures one can see that the
experimental values are reproduced quite well for the both typical
examples $-$ with enhanced $B(E2)$ values ($^{157}Gd$) and with
cutoff ($^{163}Dy$).

\section{Conclusions}

In this work we extended the dynamical symmetry group $Sp(12,R)$ of
the IVBM to the orthosymplectic one $OSp(2\Omega/12,R)$. We
introduced the fermion degrees of freedom by means of including a
particle (quasiparticle) with spin $j$ and exploiting the
corresponding reduction $SO^{F}(2\Omega)\supset SU^{F}(2)$.

Further, the basis states of the odd systems are classified by the
new dynamical symmetry (\ref{chain}) and the model Hamiltonian is
written in terms of the first and second order invariants of the
groups from the corresponding reduction chain. Hence the problem is
exactly solvable within the framework of the IVBM which, in turn,
yields a simple and straightforward application to real nuclear
systems.

We present results that were obtained through a phenomenological fit
of the models' predictions for the spectra of collective states to
the experimental data for odd$-A$ nuclei and their even-even
neighbors, used as a core for the formers. The good agreement
between the theoretical and the experimental band structures
confirms the applicability of the newly proposed dynamical symmetry
of the IVBM. The success is based on the (ortho)symplectic
structures of the model which allow the mixing of the basic
collective modes $-$rotational and vibrational ones arising from the
yrast conditions. This allows for the proper reproduction of the
high spin states of the collective bands and the correct placement
of the different band heads.

For two of the three isotopes considered, the $B(E2)$ transition
probabilities are calculated and compared with the experimental
data. The important role of the symplectic extension of the model
for the correct reproduction of the $B(E2)$ behavior, observed at
high angular momenta, is revealed.

The supersymmetry group $OSp(2\Omega/12,R)$ which is natural
generalization of the dynamical symmetry group $Sp(12,R)$ of the
IVBM could be further used to examine the correlations between the
spectroscopic properties of the neighboring even-even, odd-even and
odd-odd spectra of the neighboring nuclei and the underlying
supersymmetry which might be considered in nuclear physics as proved
experimentally \cite{NSUSY}. These investigations are the subject of
the forthcoming paper, but our preliminary results obtained in this
work already suggest the typical signatures of the nuclear
supersymmetry.

\section*{Acknowledgments}

The author is grateful to Dr. A. I. Georgieva and Dr. V. P. Garistov
for the helpful discussions. This work was supported by the Bulgarian
National Foundation for scientific research under Grant Number $\Phi
-1501$.
\\

\end{document}